\documentclass[namedreferences]{solarphysics}
\usepackage[optionalrh,solaromanenum]{spr-sola-addons} 
\usepackage{graphicx}                    
\usepackage{color}
\usepackage{breakurl}                         

\usepackage{ulem}

\begin{document}

\begin{article}

\begin{opening}

\title{Energy Definition and  Minimization in Avalanche Models for Solar Flares}


\author[addressref={Umontreal},corref,email={henri.lamarre@umontreal.ca}]{\inits{H}\fnm{Henri }\lnm{Lamarre}\orcid{0009-0001-9950-529X}}
\author[addressref={Umontreal},email={paul.charbonneau@umontreal.ca}]{\inits{P. }\fnm{Paul }\lnm{Charbonneau}\orcid{0000-0003-1618-3924}}
\author[addressref={CEA},email={antoine.strugarek@cea.fr}]{\inits{A. } \fnm{Antoine } \lnm{Strugarek}\orcid{0000-0002-9630-6463}}


\runningauthor{Lamarre, Charbonneau and Strugarek}
\runningtitle{Energy in avalanche models for solar flares}

\address[id={Umontreal}]{Physics Department, Universit\'e de Montr\'eal, CP 6128 Centre-Ville, Montr\'eal, Qc H3C-3J7, Canada}
\address[id={CEA}]{Universit\'e Paris-Saclay, Universit\'e Paris Cit\'e, CEA, CNRS, AIM, 91191, Gif-sur-Yvette, France}

\begin{abstract}

Self-organised critical avalanche models are a class of cellular automata that, despite their simplicity, can be applied to the modeling of solar (and stellar) flares and generate robust power-law distributions in event size measures. However, bridging the conceptual gap to both magnetohydrodynamics and real flare observations continues to prove challenging.
In this paper, we focus on a specific, key aspect of this endeavour, namely the definition of magnetic energy and its consequences for the model's internal dynamics and energy release statistics. 
We show that the dual requirement of releasing energy and restoring local stability demands that the instability criterion and boundary conditions be set in a manner internally consistent with a given energy definition, otherwise unphysical behavior ensues, e.g., 
negative energy release.
Working with three energy definitions previously used in the literature, we construct such internally consistent avalanche models and compare/contrast their energy release statistics. 
Using the same set of models, 
we also explore a recent proposal by Farhang et al.
(2018, 2019), namely that avalanches/flares should maximize the amount of energy released by the lattice when instabilities are triggered. This tends to produce avalanches of shorter duration but higher peak energy release, but adding up to similar total energy release.
For the three energy definition we tested, such avalanche models exhibit almost identical distributions of event size measures. Our results indicate that the key to reproduce solar-like power-law slopes in these size measures is lattice configurations in which most nodes remain relatively far from the instability threshold.
\end{abstract}


\keywords{Avalanche models - Solar Flares}

\end{opening}


\section{Introduction}\label{s:Introduction} 

Many natural physical systems exhibit energy loading and release spanning a very wide range of characteristic spatiotemporal scales. Examples
include landslides, forest fires, earthquakes, geomagnetic substorms, as well as solar
and stellar flares \citep[see, e.g.][Chapter 1]{Aschwanden11}.
In such physical systems, the buildup of energy is a slow, continuous process, while
its release is rapid and spatiotemporally intermittent,
with size measures of energy release events distributed over many orders of magnitude.

Even if the underlying governing physical laws are known, the computational modelling
of such multiscale systems is extremely arduous in practice. Direct numerical simulations
typically compromise on the range of scales modelled, by truncation at either or both the low and high ends of the scale.
These truncated scales can then be modeled via
subgrid models or
boundary conditions, respectively  \citep{shibata2011solar}. When this approach is not feasible, one must turn to simplifications at the level of the system geometry and/or governing physics.
Lattice-based models represent an extreme example of this latter approach.

Simply put, a lattice model is a network of interconnected nodes, each characterized by a state (or value) which evolves discretely in time according to an update rule
determined by the state of other nodes to which it is connected.
The so-called avalanche (or sandpile) models are a class of lattice-based models which has proven robust generators
of discrete  energy release events patterns exhibiting scale-invariant power-law statistics in their size measures.
A particularly interesting class of such avalanche models are those exhibiting
Self-Organised Criticality
\citep[hereafter SOC: see, e.g.][]{Baketal88,jensen1998self,Aschwanden13}.
This refers to systems in which
the statistical equilibrium state generating scale-invariant behavior is an attractor of the internal dynamics, and thus is reached and sustained without the need to fine tune the external driver or internal model parameters.

The application of such SOC avalanche models to solar flares was initiated by the seminal work of \cite{LuHamilton91}, and was successful in robustly reproducing the observed power-law form of the statistical distribution of event size measures
\citep[][see also \citealt{Luetal93}]{Dennis85}.
This success continues to motivate the search for a convincing
physical underpinning of the loading, instability and redistribution rules for the nodal variable defined over the lattice
\citep[][see also \citealt{Aschwanden13}, Chapter 12]{Lu95,Islikeretal98,Islikeretal00,Liuetal02,Farhangetal18}.

The majority of extant SOC avalanche models for solar and stellar
flares have adopted a Lu \& Hamilton-like modelling framework.
A nodal variable related to the magnetic field is defined on a Cartesian grid with nearest-neighbour connectivity, and evolves according to loading, stability and local redistribution rules through synchronous lattice updating
\citep[but do see, e.g.,][for examples of SOC flare models using fieldlines as dynamical elements]{Hughesetal03,MoralesCharbonneau08,Lopez-FuentesKlimchuk10}.

However designed, all such models leave
a good measure of 
arbitrariness in computing quantitative model output of the type that can be compared to observations. Some {\it sine qua non} constraints are nonetheless generally
agreed upon. Observations have established unambiguously that
flaring taps into magnetic energy of coronal magnetic structures overlaying
active regions; any redistribution rule should thus reduce the energy content of the lattice. Flare onset is generally believed to result from magnetic reconnection, itself triggered by magnetohydrodynamic (MHD) and/or plasma instabilities. In the Parker nanoflare scenario, the participating instability is associated with a threshold in the electrical current density at sites of magnetic tangential discontinuities.
Under this {\it Ansatz} the redistribution rules should decrease the local current density at the unstable lattice nodes.

These matters become of the utmost importance when using avalanche models to carry out flare
prediction \citep{Belangeretal07,StrugarekCharbonneau14,Thibeaultetal22}.
A physically well-motivated measure of energy is essential to define appropriate
redistribution rules, and, in a data assimilation and prediction context,
to match the model's energy release time series to observations, as done for example by \cite{Thibeaultetal22} with GOES X-Ray flux time series.
The definition of lattice energy, in turn, is critically dependent on the physical identification of the dynamical variable defined at lattice nodes and redistributed in the course of avalanches.

We begin in Section \ref{sec:LH} by reviewing the design of the \cite{Luetal93} avalanche model, and a plausible physical interpretation that can be placed on its components and dynamical rules. We introduce
in Section \ref{sec:Ealternate} two ``better'' definitions of lattice energy, and in Section \ref{sec:Edefdyn} we demonstrate how to design sandpile instability rules coherent with these new energy definitions, 
and investigate the resulting energy release statistics. In Section \ref{sec:EnergyOptimization}, using these consistent energy/stability definitions, we explore the behavior of the \cite{Luetal93} model under the hypothesis that
redistribution during avalanches must maximize energy release, as proposed originally by \cite{Farhangetal18}. We close in Section \ref{sec:DiscConcl}
by summarizing our findings in the context of the design of internally consistent lattice models of solar flares.

\section{The Lu \& Hamilton model\label{sec:LH}}

\subsection{A Reference Model\label{ssec:LHref}}

The avalanche model used throughout this paper is that originally designed by
\cite{LuHamilton91}, more specifically the version described in \cite{Luetal93} (hereafter L93; see also \citealt{Charbonneauetal01}).
The model is built on a Cartesian lattice with von Neumann nearest-neighbour connectivity, with the nodal variable identified with the magnetic vector potential ${\bf A}$.
For reasons to be described presently, we depart from L93 in considering a two-dimensional lattice and a single scalar component $A_{i,j}$ of the vector potential as
nodal variable, the latter producing a similar avalanching behavior as the vector nodal variable originally introduced by L93 in view of the adopted driving scheme
\citep{Robinson94}.

Following L93, the lattice is driven by sequentially
adding small increments $\delta A$ at randomly selected
lattice nodes, with $\delta A$ extracted from a uniform distribution of random deviates spanning the range $[-0.2,0.8]$,
and $A_{i,j}=0$ enforced at boundary nodes at all times. As the driving gradually builds up the nodal variable on the lattice,
its curvature is monitored by computing, at each node:
\begin{eqnarray}
\label{eq:LH93-1}
\Delta A_{i,j} = A_{i,j}-{\frac{1}{4}}\sum_{N_{1}}A_{N_{1}},
\end{eqnarray}
with the index $N_{1}$ running over the four nearest neighbors:
\begin{eqnarray}
\label{eq:LH93-1b}
 N_{1}\equiv [(i+1,j),(i-1,j),(i,j-1),(i,j+1)].
\end{eqnarray}
A node is deemed unstable when $\Delta A_{i,j}$ exceeds a preset threshold $Z_c$:

\begin{eqnarray}
\label{eq:LH93-2}
|\Delta A_{i,j}|> Z_c. 
\end{eqnarray}
When this condition is satisfied at any one node, driving stops and
a portion of the nodal variable at
node $(i,j)$ is transferred isotropically to its nearest neighbors according
to the redistribution rule:
\begin{eqnarray}
\label{eq:LH93-3}
A_{i,j}^\prime = A_{i,j}-{\frac{4}{5}}Z,\qquad A_{N_{1}}^\prime = A_{N_{1}}+{\frac{1}{5}}Z,
\qquad Z={\rm sign}(\Delta A_{i,j})Z_c,
\end{eqnarray}
where primes indicate post-redistribution values.
This redistribution rule conserves the nodal variable, and it is easily shown that it reduces the curvature measure (\ref{eq:LH93-1}) by an amount $Z_c$, which restores stability at node $(i,j)$ provided $\Delta A_{i,j}<2 Z_c$, which imposes the constraint $\delta A/Z_c<1$ on the size of the driving
increment.

While stability is thus restored at the formerly unstable node,
the redistribution process may force one or more of the nearest neighbour nodes
to exceed the instability threshold (\ref{eq:LH93-2}),
in which case the redistribution rules
(\ref{eq:LH93-3}) are applied again to newly unstable node(s), and so on,
in avalanching manner, until stability is restored across the whole lattice, at which point driving resumes.

Under these evolutionary rules
the system eventually reaches a statistically stationary self-organised critical state, characterized by scale-free avalanches of size ranging from a single node
to the whole lattice, unfolding in the outer layer of a ``sandpile'' of nodal variable of approximately parabolic shape in both lattice direction and peaking at lattice center \citep[see, e.g., Section 2 in][]{Charbonneauetal01}. 

\subsection{Physical Interpretation\label{ssec:LHphys}}

Building on LH93 (see also \citealt{Lu95b}), and inspired also by
the coronal heating scenario by nanoflares developed by
\cite{Parker88}, \cite{Strugareketal14} propose
a specific interpretative physical picture, which is also adopted in what follows.
The 2D lattice is viewed as a perpendicular section of a coronal loop taken at its apex, where flare onset is often observed
\citep[see, e.g.,][]{Tsunetaetal92,Masudaetal94}, with the magnetic field then dominated by its axial component ($B_z$, say). With the nodal variable identified with the $z$-component of the vector potential, its curl then
defines the deviation of the magnetic field from the axial direction, i.e., the twisting and braiding of magnetic-field lines about each other.
If the twist angle is small
---\cite{Parker88} estimates it at $\simeq 14^\circ$,--- then both {\bf A}
and ${\bf B}$ are dominated by their $z$-component, in which case the electrical current density ${\bf J}$ can be approximated as
\begin{equation}
\label{eq:J}
\mu_0{\bf J}=\nabla\times {\bf B}\simeq \nabla\times\nabla\times (A_z\hat{\bf z})
=-\nabla^{2}_{\perp}A_z,
\end{equation}
under the
Coulomb gauge $\nabla\cdot {\bf A}=0$, and with the Laplacian $\nabla^{2}_\perp$
defined in the cross-sectional plane, i.e.,
perpendicular to the loop axis.

Under this interpretation, the addition of an increment $\delta A$ at a node amounts to an increase of the local twist \citep{Luetal93}, and the stability
measure $\Delta A_{i,j}$ defined via Equation \ref{eq:LH93-1} can be interpreted
as a second-order centered finite difference representation of the 2D Laplacian.
If that interpretation 
is accepted\footnote{This has been common practice in attempts to bridge the gap between MHD and lattice models of flares \citep[e.g.][]{Islikeretal00,Liuetal02,Strugareketal14}. However,
lattice models are fundamentally discrete systems, and in general the state of a node should not be interpreted as a continuous variable sampled on a computational mesh; in particular, because the nodal quantity varies discretely from one node to the next,
finite difference expressions of derivatives diverge, rather than converge, as the grid spacing tends to zero. We nonetheless proceed with the interpretation of lattice model rules in terms of centered
finite difference on the lattice.},
then the instability condition (\ref{eq:LH93-2}) becomes a threshold
on the magnitude of the electrical current density,
which is physically satisfying if the energy
released by avalanches is taken through occur via magnetic reconnection triggered by plasma instabilities.

\subsection{Defining Magnetic Energy\label{ssec:LHenergy}}

A proper definition of the magnetic-energy content ($E$)
of the lattice is clearly crucial in trying to bridge the gulf between such simple avalanches
models and actual solar (or stellar) flares.
If the nodal variable is identified with the magnetic field itself, as done
originally in \cite{LuHamilton91}, then unambiguously $E\propto\sum {\bf B}^2$ over all lattice nodes.
However, under such an identification, the solenoidal constraint
$\nabla\cdot{\bf B}=0$ is not necessarily satisfied by the set of forcing and redistribution rules described in Section \ref{ssec:LHref}. 
Identifying instead the nodal variable with the magnetic-vector potential satisfies
the solenoidal constraint by
construction, but the definition of lattice energy becomes trickier. 

In our adopted physical picture, the magnetic field components in the plane of the lattice is
${\bf B}_\perp\propto \nabla\times( A_z\bf e_{z})$. With $B_z$ fixed by conservation of magnetic flux in the loop's cross-section, the contribution of $B_x,B_y$ to the magnetic energy is $(\nabla\times (A_z\bf e_{z}))^2$.
If the redistribution/reconnection simply reduces the local twist without altering the global magnetic configuration, then one would expect
$\sum A_{i,j}^2$ to be a good proxy for the lattice magnetic energy available 
for flaring\footnote{In our geometrical interpretation of the 2D lattice (Section \ref{ssec:LHphys}), the flux perpendicular to the lattice plane must be conserved.
Therefore, if the redistribution alters only the degree of local twist, set by $B_x$ and $B_y$, and not the local $B_z$, then the contribution of $B_z$ to lattice energy never changes, and the ``free energy'' of the magnetic field is associated entirely with its components in the plane of the lattice.}. 
This is the magnetic energy definition
adopted in LH93 and many subsequent works {\citep[e.g.][]{Luetal93,GeorgoulisVlahos1998,StrugarekCharbonneau14,MoralesSantos20,Thibeaultetal22}}:
\begin{equation}\label{eq:A2}
    E_{A} = \sum_{i,j}A_{i,j}^2.
\end{equation}
Here and in the forthcoming alternate expressions for lattice energy, we have omitted the usual $1/2\mu_0$ prefactor,
which amounts to rescaling energy units.

Under this definition of magnetic energy, it is a simple matter to show
that the redistribution rule (\ref{eq:LH93-3}) not only restores stability, but also reduces the lattice energy, as it should in the flaring context. Specifically, with lattice energy $\propto\sum A^{2}$, the energy content of the five nodes involved in the redistribution rule (\ref{eq:LH93-3}) drops by
\begin{eqnarray}
    \Delta E_{A} &=& \frac{4}{5}Z_{c}\left(2 |\Delta A_{i,j}| - Z_{c}\right),
    \label{eq:curvA2}
\end{eqnarray}
as detailed in appendix \ref{app:A^{2}}. Setting $\Delta A_{i,j}=Z_c$, i.e., a node just reaching the instability threshold, sets the smallest amount ($e_0$) of energy that can be released by the system:
\begin{equation}
\label{eq:LH93-5}
e_0={\frac{4}{5}}Z_c^2.
\end{equation}
From Equation \ref{eq:curvA2}, the requirement that $\Delta E_A>0$ translates into the constraint
\begin{equation}
\label{eq:Erpos}
    |\Delta A_{i,j}| > \frac{1}{2} Z_{c},
\end{equation}
which is less constraining on $|\Delta A_{i,j}|$ than the original instability criterion of the LH93 model, as per Equation \ref{eq:LH93-2}. Reducing curvature significantly below the stability threshold, i.e. hysteresis, is actually required for energy to slowly accumulate in the lattice, and to be subsequently released by scale-free avalanches \citep{Lu95b}.

\section{Alternative Energy Definitions in the L93 Model}\label{sec:Ealternate}

The most natural definition of magnetic energy is
\begin{equation}\label{eq:B2}
    E_{B} = \sum_{i,j}\textbf{B}_{i,j}^2.
\end{equation}
As argued in Section \ref{ssec:LHenergy}, in our geometrical setup only the magnetic energy associated with the magnetic-field components in the lattice plane can be tapped into to release energy. In this context, the magnetic vector potential is only directed along $z$ and is denoted ${\bf A} = A {\bf e}_z $. We set $A$ to be the
nodal variable, and the magnetic energy contribution of the
field components at node $(i,j)$ can be computed with (assuming unit grid spacing in the lattice plane):
\begin{eqnarray}
    \textbf{B}^2 &\equiv& \left(\nabla\times{\bf A} \right)^2 
    \nonumber \\ 
    &=& 
    \left( - \frac{\partial A}{\partial x}\right)^{2}+\left(\frac{\partial A}{\partial y} \right)^{2} \nonumber \\
    &=& \left(\frac{A_{i-1,j} - A_{i+1,j}}{2}\right)^2 + \left(\frac{A_{i,j-1} - A_{i,j+1}}{2}\right)^2 . \label{eq:B2-2}
\end{eqnarray}
In the context of our adopted geometrical picture (Section \ref{ssec:LHphys}), both the magnetic field and vector potential are dominated by their $z$-component. This allows an alternate definition of magnetic energy,
as recently proposed by
\cite{Farhangetal18}. Starting from the vector identity
\begin{equation}
    \nabla \cdot(\textbf{A}\times \textbf{B}) = \textbf{B}\cdot \nabla \times \textbf{A} - \textbf{A}\cdot\nabla \times \textbf{B},
\end{equation}
If ${\bf A}$ and ${\bf B}$ are parallel the LHS vanishes, so that
\begin{equation}
    \textbf{B}\cdot \textbf{B} = \textbf{A}\cdot(\nabla\times \textbf{B})=
    \mu_0\textbf{A}\cdot\textbf{J}
    \simeq \mu_0 A_z J_z,
\end{equation}
as per Amp\`ere's Law.
This provides yet another definition of energy:
\begin{equation}\label{eq:AJ}
E_{AJ} \equiv\sum_{i,j}A_{i,j} J_{i,j},
\end{equation}
with the $z$-component of the electrical current density computed via Equation \ref{eq:J} invoking again second-order centered finite differences with unit grid spacing:
\begin{equation}
\label{eq:AJ-2}
    \mu_0 J_{i,j} =
    4A_{i,j} - (A_{i-1,j} + A_{i+1,j} + A_{i,j-1} + A_{i,j-1}).
\end{equation}

Note that this third energy definition has the peculiarity that a node can potentially contribute {\it negatively} to magnetic energy, since $J_z$ can be of either sign, while $A_z$ is a positive quantity in our lattice model setup.

Equations \ref{eq:A2}, \ref{eq:B2}--\ref{eq:B2-2} and \ref{eq:AJ}--\ref{eq:AJ-2} thus offer three distinct energy definitions, which should be operationally equivalent under the assumptions underlying our geometrical/physical setup {\it and} provided one accepts again the use of centered finite differences on the lattice as a means of calculating the magnetic field and current density from the nodal vector potential.

\begin{figure}[!h]
    \title{Correlation of avalanche energies with different energy definitions}
    \centering
    \includegraphics[width = 12cm]{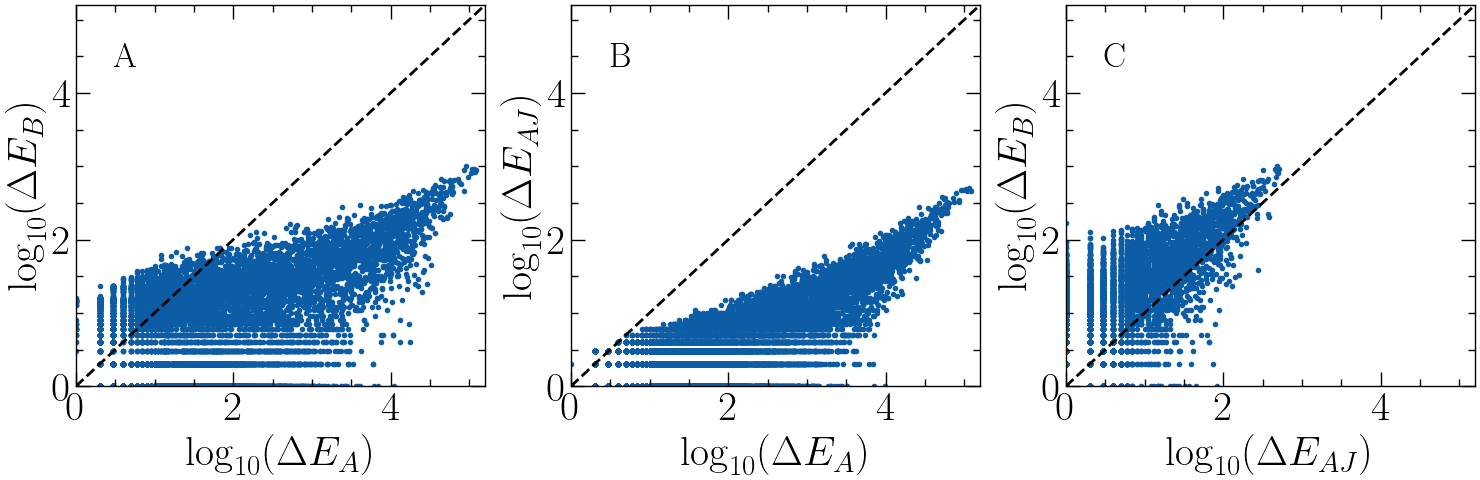}
    \caption{Comparison of the energy dissipated by avalanches generated by the LH93 model for each energy definition. The scales are logarithmic.}
    \label{fig:diffEnergies}
\end{figure}

Figure \ref{fig:diffEnergies} tests this equivalence by plotting against one another avalanche energies calculated a posteriori using our three energy definitions, working off a simulation run of the LH93 model on a 
$64\times 64$ lattice, already in the statistically stationary SOC state.
All three correlation plots show considerable scatter for the smaller avalanches, but a good correlation emerges for the larger avalanches, which is certainly encouraging.
That energy measures based on $B^{2}$ or $A J$ are systematically below those based on $A^{2}$ was to be expected, considering that under definitions (\ref{eq:B2}) and (\ref{eq:AJ}) it is possible for a node to contribute zero to lattice energy, or even a negative quantity in the case of energy definition (\ref{eq:AJ}), which is not the case with the original definition (\ref{eq:A2}).
Notably, the released energy on panel (C) aligns with
$\Delta E_B = \Delta E_{AJ}$ (dashed diagonal) quite closely, which gives confidence in the finite difference representation of derivatives used in the two alternate energy definitions introduced in this section, since these definitions should be identical under exact mathematics and under the assumptions that ${\bf B}$ is dominated by its $z$-component.

\begin{figure}[!h]
    \title{Correlation of avalanche energies with different energy definitions}
    \centering
    \includegraphics[width = 12cm]{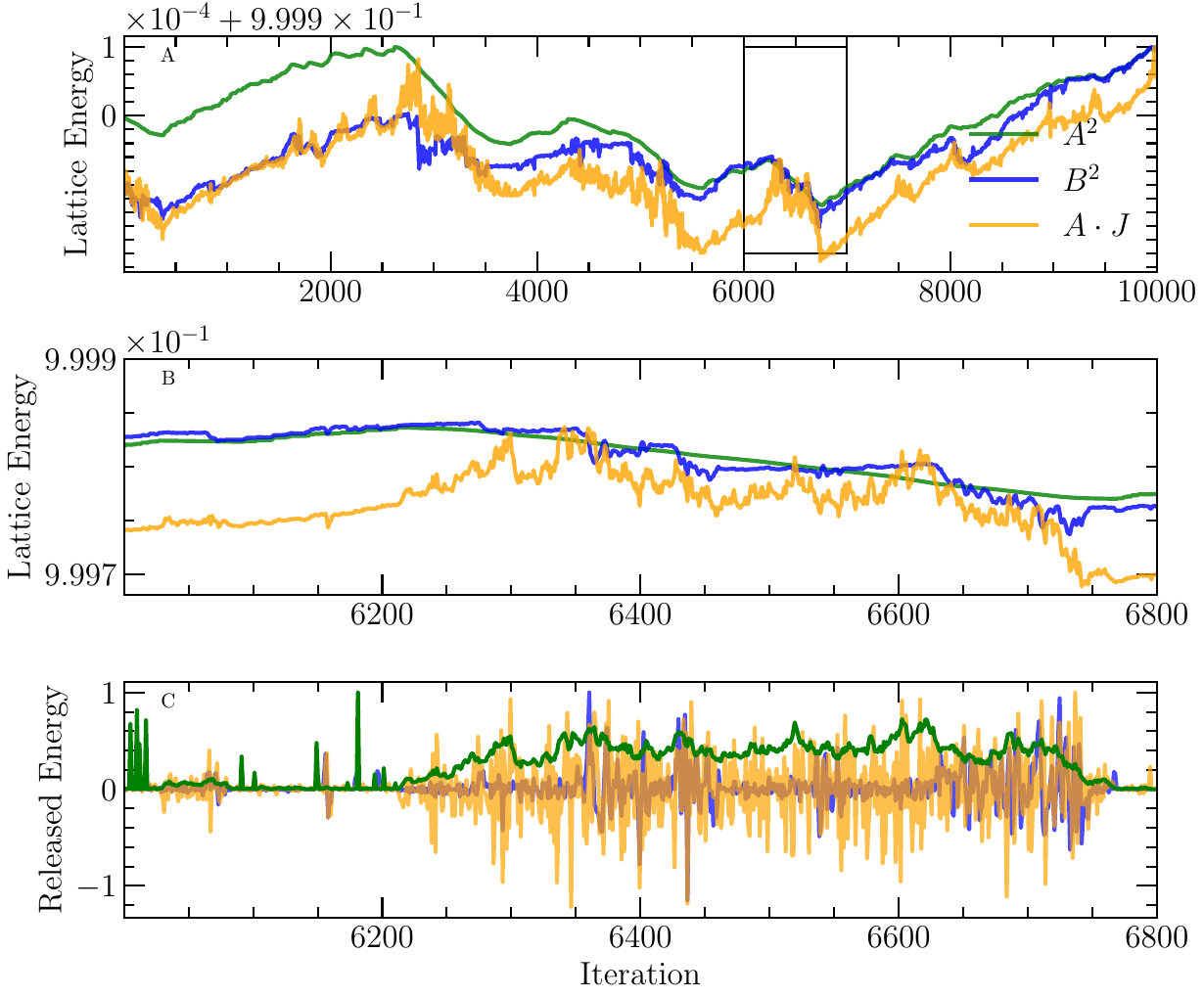}
    \caption{Panel (A): evolution of the normalized lattice energy for the LH model using the different energy definitions (\ref{eq:A2}), (\ref{eq:B2}), and (\ref{eq:AJ}), plotted in green, blue and orange respectively.
    The time series segments have each been normalized by their respective maximum values. 
    Panel (B): 1000-iteration closeup on the boxed area on panel (A).
    Panel (C): Normalized energy dissipated by the models during the time series on panel (B). The dissipated energy is the lattice energy difference for avalanching iterations and is 0 otherwise.}
    \label{fig:diffEnergiesEvolution}
\end{figure}

Unfortunately, a serious problem emerges upon closer scrutiny. 
Figure \ref{fig:diffEnergiesEvolution}A shows a $10^4$ iterations long segment
of lattice energy time series, for the same simulation used to generate the data plotted on Figure \ref{fig:diffEnergies}. The segments are color-coded according to the energy definition used, and have each been normalized to their average value over the plotted interval to facilitate visual comparison. All three time series follow the same overall trend, but the differences betray a fundamental inconsistency in the a posteriori calculations of lattice energy using the alternate definitions (\ref{eq:B2}) and (\ref{eq:AJ}). Figure \ref{fig:diffEnergiesEvolution}B,C reproduce a $10^3$ iterations
subsegment, indicated by the boxed area in panel (A), together with the time series of energy release for the three energy definitions, the latter computed as the variation in lattice energies at subsequent temporal iterations. This selected subsegment
spans a large avalanche beginning at $t\simeq 6000$. All three lattice energy time series undergo a substantial drop over the course of this large avalanche, but there are many iterations within the avalanche where the lattice energy, when computed with the $B^{2}$ or $AJ$ definition, is {\it increasing} (e.g., around $t\simeq 6300$ for $E_{B}$ and $E_{AJ}$) rather than decreasing, as it should and as it indeed does
for the $A^{2}$ energy definition (\ref{eq:A2}). This then leads to unphysical {\it negative} energy release.

This pathology arises from the fact that if the alternate energy definitions (\ref{eq:B2})--(\ref{eq:B2-2}) or (\ref{eq:AJ})--(\ref{eq:AJ-2}) are used in a model whose avalanching dynamics is governed by the stability criterion (\ref{eq:LH93-1}) and redistribution rule (\ref{eq:LH93-3}), it is no longer always the case that a redistribution that restores stability necessarily reduces lattice energy, or that a rule that decreases lattice energy necessarily restores stability. In other words, using an energy definition while using the stability criterion based on another energy definition creates cases where unstable nodes are unable to redistribute and/or avalanches release negative energy. \cite{Farhangetal18} encountered this problem and as a consequence had to introduce additional {\it ad hoc} constraints on their redistribution rules. Another possibility is to 
change either (or both) the redistribution rules or stability criterion. We opt for the latter in what follows.

\section{The Impact of Choosing an Energy Definition\label{sec:Edefdyn}}
\subsection{Constructing the Models}\label{ssec:Constructing}

In contrast to the energy definition (\ref{eq:A2}), which is purely local in the nodal variable, under the two alternate energy definitions introduced in the preceding section the calculation of energy at node $(i,j)$ involves the nodal values of nearest neighbors, as per Equations \ref{eq:B2-2} and \ref{eq:AJ-2}. As a consequence, positive energy release upon redistribution
cannot be ensured simply by adjusting the threshold value in the LH93 instability criterion (\ref{eq:LH93-2}). However, it is possible to retain a stability criterion of the general form:
\begin{equation}\label{eq:newCurv}
    \Delta A_{i,j} > \alpha Z_{c},
\end{equation}
by modifying the definition of the curvature $\Delta A_{i,j}$. As detailed in Appendix \ref{app:Rules}, these new curvature definitions involve varying numbers of next-nearest neighbors to node $(i,j)$ in Equation \ref{eq:newCurv}. These new curvature formulae, together with their corresponding energy ``quantum'' $e_0$, are compiled in Table \ref{tab:eqs}. The corresponding curvature stencils are displayed in Figure \ref{fig:Stencils} in the Appendix \ref{app:Rules}, together with the notational definition for nearest-neighbors sets $N_{1}, N_{2}$, etc.
\begin{table}[!h]  
    \begin{tabular}{c|cc|c}\hline
     ~ & $\Delta A_{i,j}$ &  $e_{0}$ & Equation \\ \hline
    $A^{2}$ &$1.6A_{i,j}-0.4 \sum N_{1}$ & $0.8(2\alpha-1) Z_{c}^2$ & \ref{eq:LH93-1}, \ \ref{eq:A.1.} \\ 
    $B^{2}$ &$1.6 A_{i,j} -0.3\sum N_{1} -0.4\sum N_{3} + 0.1\sum N_{4}$ &$0.76(2\alpha-1) Z_{c}^2$& \ref{eq:A.2.} \\ 
    $AJ$ &$4 A_{i,j} -1.6\sum N_{1} +0.4\sum N_{2} + 0.2\sum N_{3}$ &$2.24(2\alpha-1) Z_{c}^2$& \ref{eq:A.3.} \\ \hline
    \end{tabular}
    \caption{Curvature equation $\Delta A_{i,j}$ and smallest energy increment $e_{0}$ for three energy definitions.}
    \label{tab:eqs}
\end{table}

With these new energy and curvature definitions, we can build internally consistent avalanche models. However, since we are using different curvature stencils, two adjustments have been made.
\begin{itemize}
    \item First, as the new stencils probe wider than the closest lattice neighbors, the curvature computation requires additional boundary conditions at the lattice edge. For the $B^{2}$ model, we simply pad the lattice with zeros. For the $AJ$ model, we pad the lattice with negative nodal values so as to keep $J=0$ on the boundary. These choices are explained in detail in Appendix \ref{app:BC}.

    \item Second, instead of requiring that $|\Delta A_{i,j}|>Z_{c}$, we require that $\Delta A_{i,j} > Z_{c}$ as the $AJ$ model is prone to cases where neighbor nodes with opposite curvatures simultaneously redistribute which reverses the curvatures and create and alternating checkerboard patterns. This can lead to the development of very long avalanches, effectively locking the model into a non-SOC state. We note that this adjustment is only necessary for the $AJ$ model, and that applying it to the $A^{2}$ and $B^{2}$ models does not affect their statistics. For the sake of consistency, we have opted to apply this rule to all the models in what follows.
\end{itemize}

\subsection{Results}\label{ssec:ResultsEnj}

Since the models now have different boundary conditions and stability criteria, the self-organised criticality state they reach differs substantially. Examples of energy releases and lattice energy for the models are displayed in Appendix \ref{app:ModelsEvolution}. Figure \ref{fig:cone}A shows the absolute pile height in the form of 1D slices through lattice center, while panel B shows the pile height normalized to its peak central value, for models run under our three distinct energy definition and associated stability criteria (viz.~Table \ref{tab:eqs}). Both the absolute pile height and shape are very different. The absolute pile height is determined by the stability criterion, since in the statistically stationary non-avalanching state the average curvature is typically a set fraction of the stability threshold value.
The pile shape is affected by both the adopted stability criterion and associated boundary
condition. The absolute pile height is actually not important dynamically, since in the self-organised critical
state the unfolding of avalanches  is determined primarily by the distribution of curvature measures over the lattice, as plotted on Fig.~\ref{fig:cone}C.
Despite very different pile heights and shapes, the $A^{2}$ and $B^{2}$ models have closely similar
curvature distributions, while that of the $AJ$ model differs more significantly.
As we shall presently see, this translates into distinct distributions
of event size measures.

\begin{figure}[!h]
    \centering
    \includegraphics[width = 12cm]{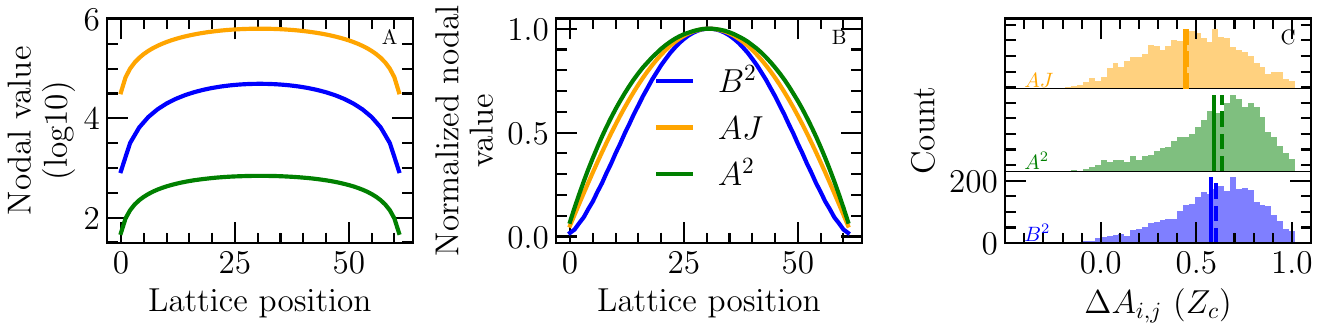}
    \caption{[A] and [B] Transversal slice across lattice center of the nodal value in models with different definitions of energies. [A] The log10 of the nodal value. [B] The nodal value divided by the maximum value along the slice. [C] Histogram of the curvature at lattice nodes for each model. For each histogram, the mean (solid line) and median (dashed line) are displayed.}
    \label{fig:cone}
\end{figure}

\begin{figure}[!h]
    \centering
    \includegraphics[width = 12cm]{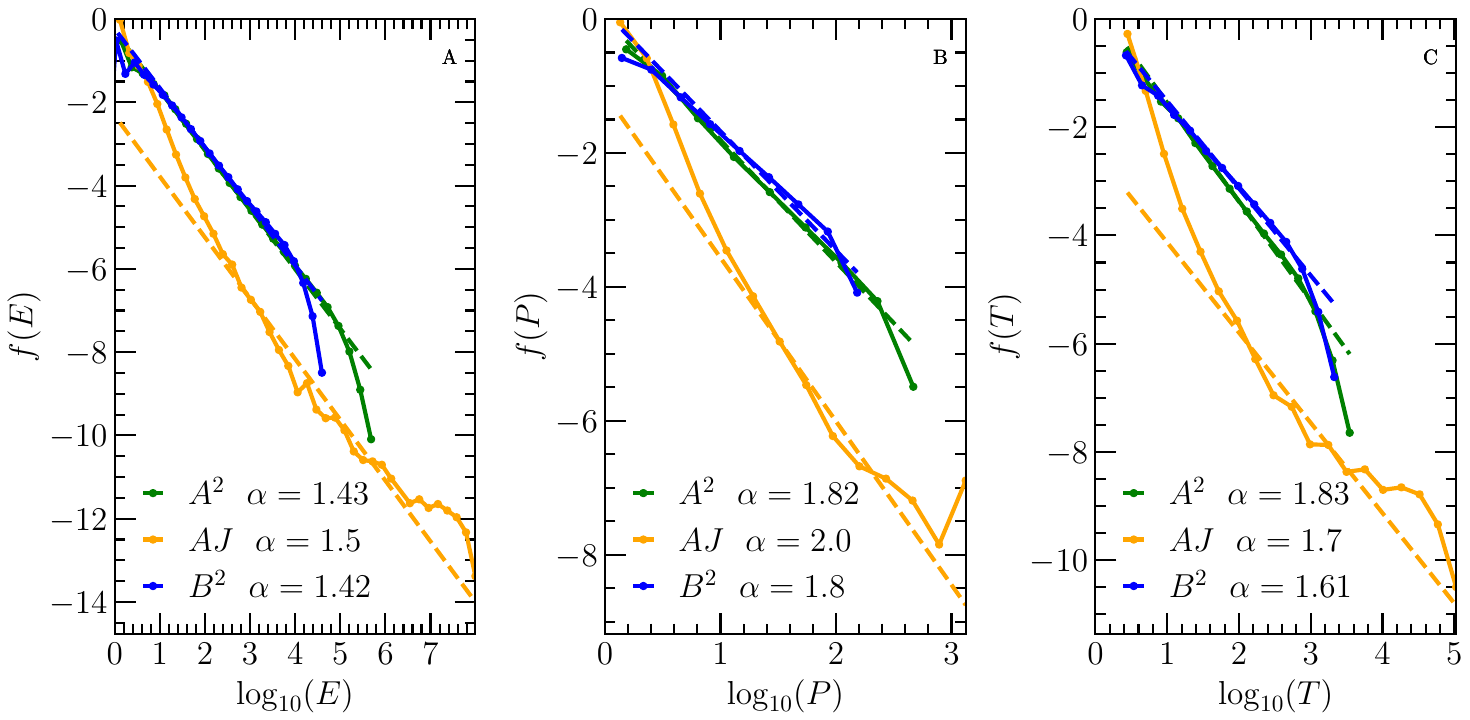}
    \caption{Probability distribution of avalanche sizes in the L93 avalanche model run under our three different energy definitions. [A] The energy dissipated by avalanches. [B] The peak energy dissipated by avalanches. [C] The duration of avalanches. For each model, a dashed line is used as a visual guide for the logarithmic slope $\alpha$ computed with the method detailed in Appendix \ref{app:Fitting}. All models are run on a 64x64 lattice with $\delta A\in[-0.2,0.8]$.}
    \label{fig:EnergyStats}
\end{figure}

Some statistical properties of avalanche have observational equivalents in solar flares: the total energy released by the avalanche ($E$), the maximal energy released in one iteration of an avalanche ($P$) and the duration of avalanches ($T$). For each model, we collect those parameters for each avalanche that occurred over $10^7$ iterations and build their frequency distributions, as shown on Figure \ref{fig:EnergyStats}. In all models, $E$ and $P$ are expressed in units of the quantum $e_0$, as listed in Table \ref{tab:eqs}. To recover the power-law exponent for each distribution, we fit the
complementary cumulative distribution function as detailed in Appendix \ref{app:Fitting}. The values for all power-law indices and associated error bars in this paper are provided in Table \ref{tab:alphas}. The corresponding power law exponents are listed in each panel.

The remarkable similarity of the event size distributions for the $A^{2}$ and $B^{2}$ models is a direct reflection of their similar distributions of
curvature measures (see Fig.~\ref{fig:cone}C).
This confirms that the choice of $A^{2}$ or $B^{2}$ as an energy definition in these models leads to equivalent
distribution of energy release, provided that the model rules are formulated in a manner consistent with the adopted energy definition.
It is also clear from figure \ref{fig:EnergyStats} that the $AJ$ model differs significantly from the other two in its statistical properties. It exhibits slightly steeper power law slopes in ($E$) and ($P$), but not as steep as the slopes reported by \cite{Farhangetal18}. This steeper power law is again consistent with the curvature distributions in figure \ref{fig:cone}: fewer lattice nodes are close to the stability threshold, which makes it harder for large avalanches to develop. The event size distributions from this model also extend to higher values
of total energies, peak energy and durations. 

\section{Energy Minimization in Redistribution}\label{sec:EnergyOptimization} 

In the solar flare context, a {\it sine qua non} requirement is that avalanches should release energy, by reducing the system's energy content. \citet{Farhangetal18} have pushed this logic one step further by designing a lattice-based avalanche model in which redistribution locally {\it minimizes} lattice energy, thus maximizing energy release. Reaching the lowest available energy state is the hallmark of closed systems relaxing to equilibrium, which can arguably be applied locally to magnetic reconnection since the associated dynamical timescales are much shorter than those characterizing the global evolution of active regions.

We follow here \cite{Farhangetal18} in introducing
redistribution rules in which the amount of nodal variable transferred to nearest-neighbors upon redistribution are no longer the same, unlike the isotropic redistribution characterizing the original LH93 model
(viz. Equation \ref{eq:LH93-3}). For simplicity, we first retain the $A^{2}$-based definition of magnetic energy (Section \ref{ssec:Opt}), and consider two formulations of energy-minimizing
redistribution: the original analytical method introduced by
\cite{Farhangetal18} (hereafter F18), and a variation based on a Monte Carlo approach (MC). In both cases redistribution is restricted to the four immediate
nearest-neighbors, as in the original LH93 model. Then, we extend the MC approach to the two other definitions of energy considered in this work ($AJ$ and $B^{2}$, Section \ref{sec:Ealternate})

\subsection{Analytical Maximization of the Energy Release}\label{ssec:Opt}

The energy-minimizing, anisotropic redistribution rule introduced by \cite{Farhangetal18} is defined as:
\begin{eqnarray}
    A_{i,j}^\prime &=& A_{i,j} - {\frac{4 Z_{c}}{5}}\nonumber, \\
    A_{i+1,j}^\prime &=& A_{i+1,j} + {\frac{r_{1}}{x+a}}{\frac{Z_{c}}{5}},\nonumber \\
    A_{i-1,j}^\prime &=& A_{i-1,j} + {\frac{r_{2}}{x+a}}{\frac{Z_{c}}{5}},\nonumber \\
    A_{i,j+1}^\prime &=& A_{i,j+1} + {\frac{r_{3}}{x+a}}{\frac{Z_{c}}{5}},\nonumber \\
    A_{i,j-1}^\prime &=& A_{i,j-1} + {\frac{x}{x+a}}{\frac{Z_{c}}{5}},
\label{eq:opt1}
\end{eqnarray}
where $a=r_1+r_2+r_3$ and
$r_{1}$,$r_{2}$,$r_{3}$ are random numbers extracted from a uniform distribution $\in [0,1]$, controlling the amount of nodal variable transferred to three of the nearest-neighbour nodes.
The idea is then to choose a value
for $x$, which sets the quantity of nodal variable transferred to the fourth nearest-neighbour node, in a manner such as to minimize lattice energy after the redistribution. In Equations \ref{eq:opt1} this fourth node was chosen as $(i,j-1)$, but in practice
the optimization direction must be chosen randomly at each redistribution, in order to ensure global isotropy in avalanching behavior. The calculation of this optimal $x$ is detailed in Appendix \ref{app:Analytical} (see Equation \ref{eq:x_optimal}), in the context of the $A^{2}$ definition of energy.

Following the methodology of \citet{Farhangetal18}, the optimal $x$ is obtained following a variational approach. Nevertheless, this approach leads to some situations that need to be treated with care. First, the variational approach only guarantees that $x$ is an extremum value, but not necessarily always a maximum. 
Second, if the optimal $x$ is negative, situations where $x\approx -a$ will destabilize the system as an arbitrarily large amount of nodal value can be distributed to the neighbors. When such situations occur, we generate new sets of $r_k$ in a new random direction each time until $x>0$ and is an actual maximum. Typically, up to 20 trials are required to achieve this.

\subsection{Monte Carlo Maximization of the Energy Release}\label{ssec:MC}

A Monte Carlo approach to lattice energy minimization can also be designed based on the anisotropic redistribution rule introduced in \cite{Strugareketal14}:
\begin{eqnarray}
\label{eq:Opt2}
    A_{i,j}^\prime &=& A_{i,j} - \frac{4 Z_{c}}{5},\\
    A_{N_{1}}^\prime &=& A_{N_{1}} + \frac{r_{k}}{a}\frac{4 Z_{c}}{5},\qquad k=1,2,3,4.
\end{eqnarray}%
with $a=\sum r_{k}$ and the four $r_k$'s
$\in[0,1]$ are again uniformly distributed random numbers. The idea is then to generate $N_{\rm MC}$ sets of $r_k$'s, calculate for each the energy that would be released upon redistribution, and retain the set member that leads to the lowest post-redistribution lattice energy (i.e. the largest energy release). Note that even in the $N_{\rm MC}\to\infty$ limit this approach does {\it not} become identical
to the \cite{Farhangetal18}
scheme of Section \ref{ssec:Opt}, since in this latter case
only the amount $x$ of nodal variable transferred in the selected optimization direction is varied to achieve minimization of lattice energy in a given redistribution event; whereas in the Monte Carlo scheme all four nearest-neighbour increments are
reset randomly at each of the $N_{\rm MC}$ minimization trial. As a consequence, the \cite{Farhangetal18} model tends to generate redistributions that are more strongly anisotropic, and does not minimize lattice energy as much as the
Monte Carlo scheme can, already at $N_{\rm MC}=20$. 

\subsection{Comparison of the Optimized Models}\label{ssec:MCvsF18}

We now compare and contrast simulation runs carried out under both energy minimization schemes, to a ``standard'' LH93 run using the isotropic redistribution (\ref{eq:LH93-3}). All simulations are carried out on a $64\times 64$ lattice, with $Z_c=1$ and forcing amplitude $\delta A\in[-0.2,0.8]$, under the original energy definition (\ref{eq:A2}).

\begin{figure}[!h]
    \centering
    \includegraphics[width=12cm]{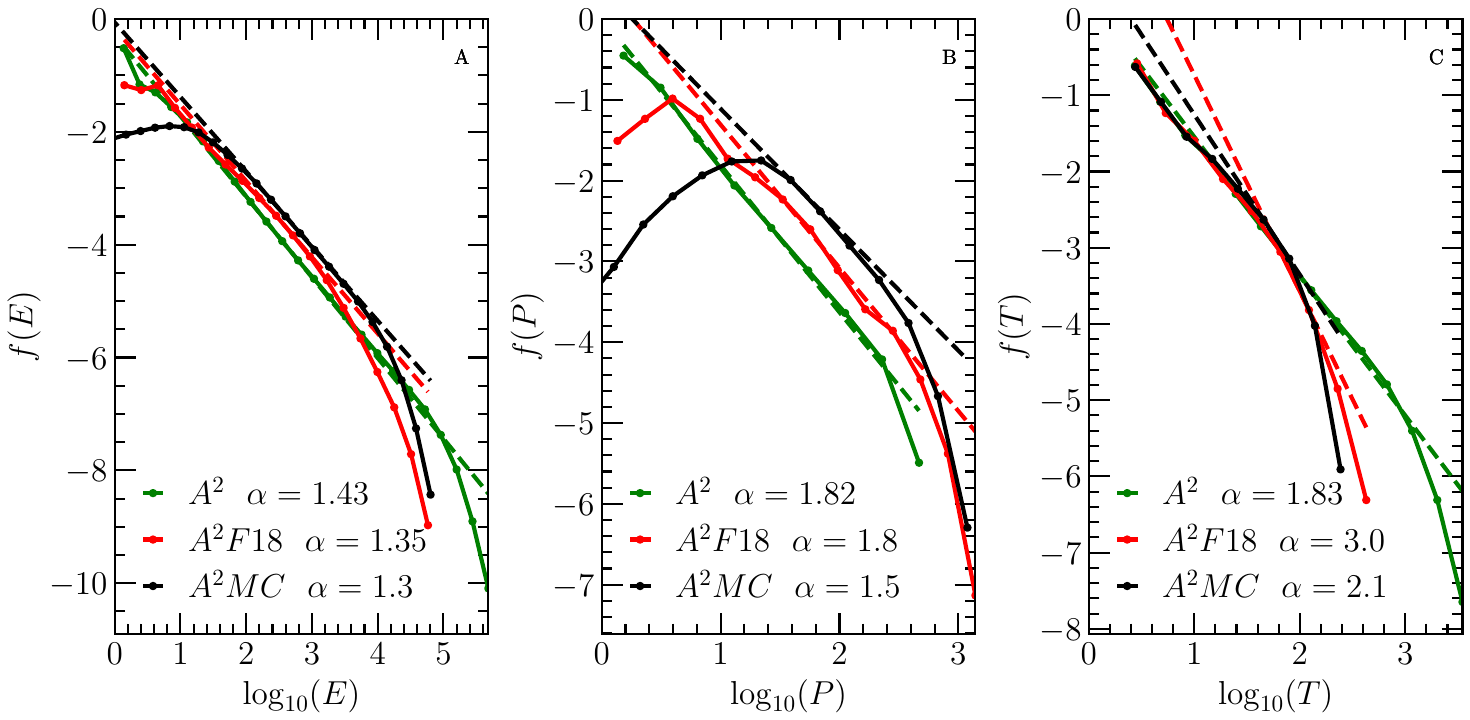}
    \caption{Probability distribution of avalanche sizes in the L93 avalanche model with 64x64 size and $A^{2}$ definition of energy. LH is the standard non optimized model. F18 is the analytical optimized model and MC is the Monte-Carlo optimized model. For each model, a dashed line is used as a visual guide for the logarithmic slope $\alpha$ computed with the method detailed in Appendix C. All models have $\delta A\in[-0.2,0.8]$ and have a fixed stability threshold and restriction for the curvature $\Delta>Z_{c}$.}
    \label{fig:OptStats}
\end{figure}

Interestingly, the power-law indices characterizing avalanche
size measures are similar between all three models. However, as shown in Figure \ref{fig:OptStats}, the  ranges of frequency distributions for these size measures are altered by energy minimization. As was to be expected, the peak energy release (panel B) is larger in both energy minimizing models, and avalanche durations (panel C) are markedly reduced, a consequence of lattice energy dropping more rapidly during avalanches developing under either energy minimization schemes. These trends are slightly more pronounced with the Monte Carlo-based minimization, which is to be expected since it samples a broader range of potential post-redistribution lattice states. It is interesting to note that neither of the optimized models are able to produce avalanches with total energy as large as the LH model (panel A). This occurs because under either energy minimization schemes, even small avalanches are more efficient at reducing lattice energy, making it more difficult for energy to accumulate in the lattice and give rise, upon destabilization, to very large avalanches. Finally, note that the MC model is so efficient at releasing energy that even small avalanches of only one or two nodes usually release a relatively large amount of energy, i.e., many times the quantum $e_0$. This depletes the distribution of energy released at small values, leading to the break of scale invariance at low energy, characterizing the MC model on panels A and B.

\subsection{Optimization Under the Different Energy Definitions}\label{ssec:MCopt}

\begin{figure}[!h]
    \centering
    \includegraphics[width=12cm]{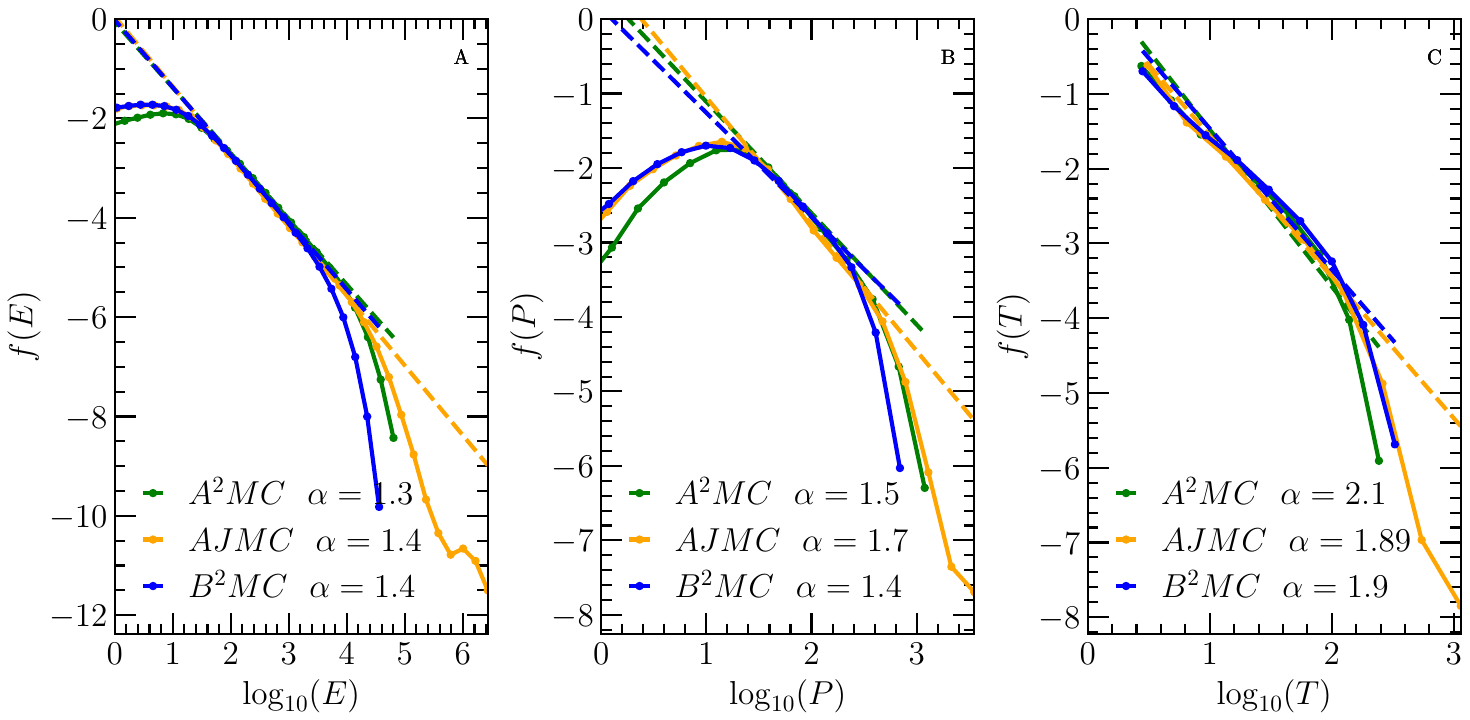}
  
\caption{Statistical properties of three Monte Carlo optimized models with $64\times 64$ size for the $A^{2}$, $B^{2}$ and $AJ$ definition of energy. For each model, a dashed line is used as a visual guide for the logarithmic slope $\alpha$ computed with the method detailed in Appendix C. All models have $\delta A\in[-0.2,0.8]$ and have a fixed stability threshold and restriction for the curvature $\Delta>Z_{c}$.}
    \label{fig:MCopt}
\end{figure}

We now apply the MC minimization approach to models constructed under our three energy definitions. The statistical properties of those models are shown in Figure \ref{fig:MCopt}. 
Comparing to Figure \ref{fig:EnergyStats}, it is clear that within the MC framework, the differences between the event size distributions under the three different energy definition models are much reduced.
This suggests that the steeper power-law indices, as first reported in \cite{Farhangetal18}, are associated with the $AJ$ energy definition, rather than with the
energy release maximization procedure.

Figure \ref{fig:MCopt} also suggests that redistribution rules maximizing energy release, when pushed to their limit, do succeed in releasing as much energy as can possibly be released by the lattice, under any plausible definition of lattice energy.
This would be akin the free energy of coronal magnetic structures, corresponding to the magnetic energy above that of a potential magnetic field satisfying the same boundary conditions \citep[see, e.g.,][]{Aly91}. 

\section{Discussion and Conclusion}\label{sec:DiscConcl}

The primary appeal of self-organised critical avalanche models remains their
robust reproduction of the power-law form of the statistics of event size measures.
However, going beyond the power-law form towards quantitative comparison of model predictions
to observations requires an unambiguous definition of lattice and avalanche energies, as
an essential step in bridging the gap between such simple statistical models, the magnetohydrodynamics of solar flares, and actual flare observations.

In this paper, working with a two-dimensional
variation on the original \cite{Luetal93} avalanche model,
we have investigated this question under three physically plausible
definitions of lattice energy,
all expressed in term of a nodal variable identified with a magnetic vector potential.
The first equates lattice energy to the sum of the squared vector potential; the second
computes a magnetic field ${\bf B}=\nabla\times {\bf A}$ via finite difference evaluation
of the curl operator on the lattice, and then equates energy to $B^{2}$ summed over all lattice nodes. The third is the alternate energy magnetic definition introduced
by \cite{Farhangetal18}, which computes energy as ${\bf A} \cdot {\bf J}$ summed over the lattice, ${\bf J}$ being the electric current density, again computed through finite differences
over the lattice.
Under the geometrical setup and physical interpretation of our 2D lattice and exact mathematics,
these second and  third definitions are in principle identical.

A {\it sine qua non} requirement of SOC avalanche models is that once a node exceeds the instability threshold,
redistribution must (1) locally restore stability at that node,
and (2) release lattice energy \citep[see, e.g.,][]{Lu95b}.
We have shown that in order for both of these conditions to be systematically satisfied,
the choice of a specific energy definition must be accompanied
by a consistent definition of redistribution rules and/or stability criterion accompanied by specifying boundary conditions.
Failing to do so results in avalanching nodes sometimes
releasing ``negative'' energy and/or remaining unstable after redistribution, with significant consequences on the dynamical behavior and pattern of energy release. This ends up forcing the introduction of {\it ad hoc} and physically dubious additional procedures to yield a functional model.

Running the \cite{Luetal93} under the these three energy definitions and
appropriately defined stability criteria yields avalanche statistics,
i.e. distributions of total avalanche energy, duration and peak energy release,
that are almost indistinguishable in the case of the $A^{2}$ and $B^{2}$
energy definitions. This vindicates, a posteriori, the common use of the summed squared nodal variable as a measure of lattice energy \citep[e.g.][]{Luetal93,GeorgoulisVlahos1998,StrugarekCharbonneau14,MoralesSantos20,Thibeaultetal22}.

We have also explored a physically appealing proposal, also put forth by \cite{Farhangetal18}, namely that
redistribution should not just release lattice energy, but in fact {\it maximize} the amount
of magnetic energy released during a redistribution.
We have investigated two implementations of this idea, the first essentially identical to the procedure introduced by \cite{Farhangetal18},
the other based on a Monte Carlo optimisation scheme.
The latter is more demanding computationally, but turns out to achieve higher levels
of energy release than the original optimization scheme of \cite{Farhangetal18}. Indeed,
the Monte Carlo scheme is so efficient that it exhibits a pronounced deficit of single-node
avalanches of very small energies, causing a break of scale invariance at these energies.
Excluding these smallest avalanches, under such
energy maximization schemes, avalanches are typically more intense (higher
peak energy release), of shorter duration,
but release similar total amounts of energy.

An appealing property of the energy minimizing avalanche models first proposed by
\cite{Farhangetal18,Farhangetal19} is their ability to generate steeper power laws in the size measures of energy release events, in better agreement with current observational inferences
\citep[see, e.g.,][]{aschwanden2002nanoflare, joulin2016energetic, vilangot2020power}. Working under their $AJ$ energy definition within the classical L93 model, i.e., without maximizing energy release, we recover slightly steeper
power-law slopes (viz. Figure \ref{fig:EnergyStats}) albeit not as steep as the one reported by \cite{Farhangetal18}. This difference likely stems from our implementation of this energy definition, which differs 
significantly, especially with regard to the stability criterion (Section \ref{ssec:Constructing}). 
However, applying the two schemes for energy maximization to the L93 model yields similar power-law slopes as non-energy-maximizing isotropic redistribution (see Figure \ref{fig:OptStats}). Moreover, the modelling results presented in Section \ref{ssec:MCopt} also indicate that even under the $AJ$ energy definition the steeper power-law slopes of Figure \ref{fig:EnergyStats} revert to those characterizing the other energy definitions upon imposing the strong minimization achieved by the Monte Carlo optimization scheme (cf. Figure \ref{fig:MCopt}). 
One can thus conclude that steeper power-law slopes in event size measures do not directly result from the maximization of energy release. Our modelling results suggest instead that the key is a broad curvature distribution, characterized by relatively few lattice nodes being very close to the instability threshold (viz. Figure \ref{fig:cone}C).
The \cite{Farhangetal18} model happens to achieve this through the combination of their energy definition and specific choices of update rules.

In general, closed systems near equilibrium are expected to minimize their energy in seeking equilibrium; however, a flaring active region is not a closed system, is likely far from equilibrium, and the reconfiguration of the magnetic field is subjected to topological constraints
posed by MHD invariants such as magnetic helicity.
Whether or not a flaring site within an active region can really relax to a locally minimal energy state (consistent with boundary conditions) is an extremely interesting physical question, which remains entirely open at this juncture.

\begin{appendix}
\section{Curvature Definitions}\label{app:Rules}

We present here the detailed calculations leading to the curvature definitions and energy quantas presented in Table \ref{tab:eqs}, Section \ref{sec:Edefdyn}. The goal is to
ensure the decrease of lattice energy during redistribution. We opted to retain the nearest-neighbour redistribution rule (\ref{eq:LH93-3}) and a stability criterion of the general form (\ref{eq:newCurv}), but
alter the definition of curvature $\Delta A_{i,j}$ according to the energy definition adopted.

\subsection{Curvature Stencils and Boundary Conditions\label{app:BC}}

\begin{figure}[!h]
    \centering
    \includegraphics[width=10cm]{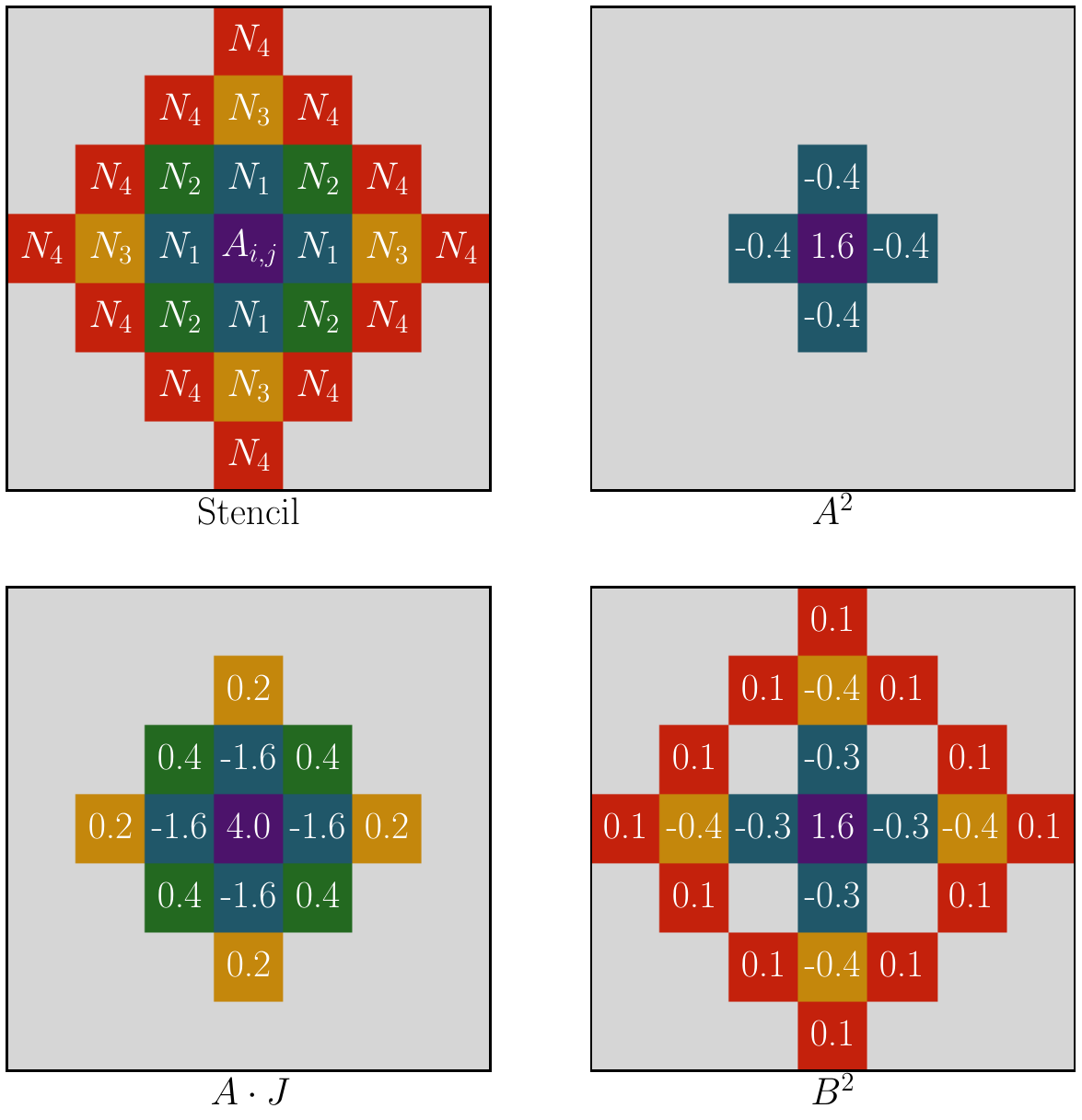}
    \caption{Visualisation of the different curvature stencils. Each of them applies to a curvature inequation of the form $\Delta A_{i,j} > Z_{c}$ where $\Delta A_{i,j}$ is a weighted sum over all cells, with weights given in the stencils. Empty cells are not used for calculation of the curvature $\Delta A_{i,j}$.}
    \label{fig:Stencils}
\end{figure}

The new curvature expressions for the two alternate energy definitions introduced in Section \ref{sec:Ealternate} end up involving more nodal neighbours than under the original energy definition $(\ref{eq:A2})$ used in L93. Figure \ref{fig:Stencils} depicts the stencils and associated notation used in this appendix to describe the new curvature formulae.

As with the original L93 model, under all energy definitions the nodal variable $A_{i,j}$ is reset to zero at every (avalanching) iteration on all boundary nodes. In view of the driving and use of 
conservative redistribution rules, this is a {\it sine qua non} condition for reaching
a statistically stationary state. This is the only boundary condition required in the $A^{2}$ model. 
However, the wider stencils used along with the $B^{2}$ and $AJ$ energy definitions require additional 
boundary conditions. 

For the $B^{2}$ stencil, one $N_3$ and three $N_4$ neighbours lie outside the lattice
when applying the curvature stencil to an interior node adjacent to the lattice boundary. Additional $N_3$ and $N_4$ are involved when considering the four interior corner nodes. We opted to simply pad the lattice with two layers of ghost nodes, on which $A=0$ is enforced at all times.

In the case of the $AJ$ stencil, one $N_3$ neighbour can lie outside the lattice, with two additional $N_3$ nodes in the case of internal corner nodes. However, simply padding the lattice with a layer of $A=0$ ghost nodes leads to $J$ changing sign when moving from the edge of the lattice to the center, while $A$ remains positive. Under
the $AJ$ energy definition (\ref{eq:AJ}), this produces an annular region adjacent to the lattice boundaries where nodes contribute {\it negatively} to lattice energy. 
This is clearly an unphysical situation, which moreover ends up affecting significantly
the statistics of event size measures in this model.
Motivated 
by our physical/geometrical picture (viz. Section \ref{ssec:LHphys}), we require instead
that the current $J$ be zero on the lattice boundary, 
as required in the vacuum exterior to
a coronal loop. This is achieved setting the value of $A$ at each ghost nodes equal to the negative of the value
of its corresponding interior node, as per Equation \ref{eq:AJ-2}. All interior nodes then contribute positively to lattice energy.

Under these boundary conditions, edge nodes have zero energy in the $A^{2}$ and $AJ$ models. This condition is not striclty achieved in the $B^{2}$ models, but their contribution to lattice energy remains insignificant.

\subsection{Stability Rule for $A^{2}$ Model.}\label{app:A^{2}}

Consider a single node $(i,j)$ exceeding the instability threshold and redistributing to its nearest neighbors according to the rule (\ref{eq:LH93-3}). Under the energy definition (\ref{eq:A2}), the change in lattice energy denoted $\Delta E_A$ is
\begin{eqnarray}\label{eq:A.1.}
    \Delta E_A &=& \left(A_{i,j}^2+\sum A_{N_{1}}^2\right)-\left( (A_{i,j}-\frac{4}{5}Z_c)^2+\sum (A_{N_{1}}+\frac{1}{5}Z_c)^2\right) \nonumber \\
    &=&\frac{4}{5}Z_{c}\left(2A_{i,j}-\frac{1}{2}\left(A_{i+1,j}+A_{i-1,j}+A_{i,j+1}+A_{i,j-1}\right)- Z_{c}\right) \nonumber \\
    &=&\frac{4}{5}Z_{c}\left(2\Delta A_{i,j}- Z_{c}\right),
\end{eqnarray}
%
%
the last step making use of
the curvature definition (\ref{eq:LH93-1}) and $N_{1}$ corresponding to the four neighbors depicted in Figure \ref{fig:Stencils}B.
A positive energy release $\Delta E_A > 0$ thus leads to the constraint 
(\ref{eq:Erpos}):
\begin{equation}
    \Delta A_{i,j} > \frac{1}{2} Z_{c}.
\end{equation}
This constraint can be generalized in the form
\begin{equation}
    \Delta A_{i,j} > \alpha Z_{c},
\end{equation}
with $\alpha > 1/2$. The minimal energy release  ``quantum'' $e_0$ is then obtained when  $\Delta A_{i,j}=\alpha Z_c$ and is given by
\begin{equation}
    \label{eq:e0A2}
    e_0=\frac{4}{5}(2\alpha-1)Z_c^2.
\end{equation}
This generic formulation reduces to Equation \ref{eq:LH93-5} for $\alpha=1$.

The redistribution must also restore stability. Consider a situation where node $(i,j)$ is barely below the instability threshold, i.e.: 
\begin{equation}
\label{eq:dum1}
    A_{i,j} - \frac{1}{4}\sum_{N_{1}}A_{N_{1}} = \alpha Z_{c} - \epsilon,
\end{equation}
with $\epsilon\ll \alpha Z_c$. Upon adding an increment $\delta A$ ($>\epsilon$) to $A_{i,j}$, the instability threshold is exceeded and redistribution takes place. Post-redistribution, the curvature should be reduced below
$\alpha Z_c$:
\begin{eqnarray}
    (A_{i,j}+\delta A - \frac{4}{5}Z_{c}) - \frac{1}{4}\left(\sum_{N_{1}}A_{N_{1}} + \frac{1}{5}Z_{c}\right) &<& \alpha Z_{c},\nonumber \\
    A_{i,j}+\delta A - \frac{4}{5}Z_{c} - \frac{1}{4}\sum_{N_{1}}A_{N_{1}} - \frac{1}{5}Z_{c} &<& \alpha Z_{c},\nonumber \\
    A_{i,j} - \frac{1}{4}\sum_{N_{1}}A_{N_{1}} +\delta A - Z_{c} &<& \alpha Z_{c},\nonumber \\
    \alpha Z_c-\epsilon +\delta A - Z_{c} &<& \alpha Z_{c},
\end{eqnarray}
where (\ref{eq:dum1}) was used in the last step.
Even with $\epsilon\ll 1$, stability is always restored provided the increments $\delta A$ is small enough to satisfy:
\begin{equation}
    \delta A < Z_{c}.
\end{equation}
\subsection{Stability Rule for $B^{2}$ Model.}\label{app:B^{2}}

Similarly, for the ${\bf B}^2$ energy definition, consider a single node $(i,j)$ exceeding the instability threshold and redistributing to its nearest neighbors according to the rule (\ref{eq:LH93-3}). Under the energy definition (\ref{eq:B2}), (\ref{eq:B2-2}), the change in lattice energy denoted $\Delta E_B$ is
\begin{eqnarray}
\Delta E_B=&\left(B_{i,j}^2+\sum B_{N_{1}}^2+\sum B_{N_{2}}^2+B_{N_{3}}^2\right)-\\&\left(B_{i,j}'^2+\sum B_{N_{1}}'^2+\sum B_{N_{2}}'^2+B_{N_{3}}'^2\right), \nonumber
\end{eqnarray}
where $N_{1}$, $N_{3}$ and $N_{4}$ are the three groups of neighbors, as displayed in Figure \ref{fig:Stencils}D.  We can expand this expression by detailing the energy evolution over single nodes. First we have
\begin{eqnarray}
    B_{i,j}^2 - B_{i,j}'^2 &=&
    (A_{i-1,j} - A_{i+1,j})^2 + (A_{i,j-1} - A_{i,j+1})^2 -\nonumber\\ &&(A_{i-1,j} + \frac{1}{5}Z_{c} - A_{i+1,j} - \frac{1}{5}Z_{c})^2 -\nonumber\\ &&(A_{i,j-1} + \frac{1}{5}Z_{c} - A_{i,j+1} - \frac{1}{5}Z_{c})^2\nonumber\\
     &=& 0
\end{eqnarray}
Then, for the energy of a $B_{N_{1}}$ node, we show an example of derivation for $B_{i,j+1}$ (the computation is similar for the other nodes):
\begin{eqnarray}
    B_{i,j+1}^2 - B_{i,j+1}'^2 &=&
    (A_{i-1,j+1} - A_{i+1,j+1})^2 + (A_{i,j} - A_{i,j+2})^2 -\nonumber\\ &&(A_{i-1,j+1} - A_{i+1,j+1})^2 - (A_{i,j} - \frac{4}{5}Z_{c} - A_{i,j+2})^2 \nonumber\\
     &=&Z_{c}\left(1.6A_{i, j+2} - 1.6 A_{i,j} - 0.64 Z_{c}\right).
\end{eqnarray}
Similarly, for the energy of a $B_{N_{2}}$ node we obtain
\begin{eqnarray}
    B_{i+1,j+1}^2 - B_{i+1,j+1}'^2 &=&
    (A_{i,j+1} - A_{i+2,j+1})^2 + (A_{i+1,j} - A_{i+1,j+2})^2 \nonumber\\&&- (A_{i,j+1} - \frac{1}{5}Z_{c} - A_{i+2,j+1})^2 -\nonumber\\&& (A_{i+1,j} - \frac{1}{5}Z_{c} - A_{i+1,j+2})^2 \nonumber\\
      &=&Z_{c}\left(0.4(A_{i+1, j+2} + A_{i+1, j}+A_{i+2, j+1}+A_{i, j+1}) \right.\nonumber\\&&- \left.0.08Z_{c}\right).
\end{eqnarray}
Finally, for the energy of a $B_{N_{3}}$ node:
\begin{eqnarray}
    B_{i,j+2}^2 - B_{i,j+2}'^2 &=&
    (A_{i-1,j+2} - A_{i+1,j+2})^2 + (A_{i,j} - A_{i,j+3})^2 \nonumber\\&&- (A_{i-1,j+2} - A_{i+1,j+2})^2 - (A_{i,j+1} - \frac{4}{5}Z_{c} - A_{i,j+3})^2 \nonumber\\
     &=&Z_{c}\left(0.4A_{i, j+3} - 0.4 A_{i,j+1} - 0.4 Z_{c}\right).
\end{eqnarray}
Computing the energy difference in this manner for each stencil node and assembling the results, we obtain:
\begin{eqnarray}
    \Delta E_B &=& 0.76 Z_{c}\left(\frac{1}{0.76}\left[1.6 A_{i,j} -0.3\sum A_{N_{1}} -0.4\sum A_{N_{3}} + 0.1\sum A_{N_{4}}\right] - Z_{c}  \right) \nonumber \\
    &=& 0.76Z_{c}\left(2 \Delta_{B}A_{i,j} - Z_{c} \right),
\end{eqnarray}
where we have introduced $\Delta_B A_{i,j}$ defined as 
\begin{equation}\label{eq:A.2.}
    \Delta_{B} A_{i,j} = \frac{1}{1.52}\left[1.6 A_{i,j} -0.3\sum A_{N_{1}} -0.4\sum A_{N_{3}} + 0.1\sum A_{N_{4}}\right].
\end{equation}

Similarly to case $A^{2}$ (Appendix \ref{app:A^{2}}), a positive energy release $\Delta E_B > 0$ can be expressed as 
\begin{equation}
    \Delta_B A_{i,j} > \frac{1}{2} Z_{c}.
\end{equation}
We can, again, generalize this criterion to $\Delta_{B} A_{i,j} > \alpha Z_{c}$, with $\alpha > 1/2$. When $\alpha=1$, we recover the energy release quantum
\begin{equation}\label{eq:e0B2}
    e_{0} = 0.76Z_{c}\left(2 Z_{c}- Z_{c}\right) = 0.76Z_{c}^2.
\end{equation}
The redistribution must also restore stability. Consider a situation where a node $(i,j)$ is barely below the instability threshold, i.e.:
\begin{equation}\label{eq:dum2}
    \frac{1}{1.52}\left[1.6 A_{i,j} -0.3\sum A_{N_{1}} -0.4\sum A_{N_{3}} + 0.1\sum A_{N_{4}}\right] = \alpha Z_{c} - \epsilon
\end{equation}
Consider now a node receiving an driving increment $\delta A$, and subsequently redistributing. We require that the new curvature is below the instability threshold:
\begin{eqnarray}
    \frac{1}{1.52}\left[1.6 \left(A_{i,j}+\delta A -\frac{4}{5}Z_{c}\right)-0.3\sum \left(A_{N_{1}} + \frac{1}{5}Z_{c}\right) -\right.\nonumber\\
    \left.0.4\sum A_{N_{3}} + 0.1\sum A_{N_{4}}\right] < \alpha Z_{c}
\end{eqnarray}
\begin{equation}
    \alpha Z_{c}-\epsilon+ \frac{1}{1.52}\left(1.6\delta A - 0.3\frac{4}{5}Z_{c} - 1.6\frac{4}{5}Z_{c}\right)< \alpha Z_{c}
\end{equation}
\begin{equation}
    -\epsilon+ \delta A < 0.95Z_{c}, 
\end{equation}%
where (\ref{eq:dum2}) was used in the last step.
Even with $\epsilon\ll 1$, stability is always restored provided the increments $\delta A$ is small enough to satisfy:
\begin{equation}
    \delta A < 0.95 Z_{c}.
\end{equation}

\subsection{Stability Rule for $AJ$ Model.}\label{app:AJ} 

Moving on to the  $AJ$ energy definition, consider again a single node $(i,j)$ exceeding the instability threshold and redistributing to its nearest neighbors according to the rule (\ref{eq:LH93-3}). Under the energy definition (\ref{eq:AJ}), (\ref{eq:AJ-2}), the change in lattice energy, denoted $\Delta E_{AJ}$, is
\begin{eqnarray}
\Delta E_{AJ} &=& \left(A_{i,j}J_{i,j}+\sum A_{N_{1}}J_{N_{1}}+\sum A_{N_{2}}J_{N_{2}}+\sum A_{N_{3}}J_{N_{3}}\right)\nonumber\\ &&- \left(A'_{i,j}J'_{i,j}+\sum A'_{N_{1}}J'_{N_{1}}+\sum A'_{N_{2}}J'_{N_{2}}+\sum A'_{N_{3}}J'_{N_{3}}\right)~,\nonumber .
\end{eqnarray}
where $N_{1}$, $N_{2}$ and $N_{3}$ are the neighbors as displayed in figure \ref{fig:Stencils}C.  
Following the same methodology as in \ref{app:B^{2}}, we break down the various parts of the expression of $\Delta E_{AJ}$ to obtain  
\begin{eqnarray}
    A_{i,j}J_{i,j} - A'_{i,j}J'_{i,j} &=&
    \frac{1}{2}A_{i,j}\left(4A_{i,j} - \sum A_{N_{1}}\right) \nonumber\\&&-\frac{1}{2}\left(A_{i,j} - \frac{4}{5}Z_{c}\right)\left(4\left(A_{i,j} - \frac{4}{5}Z_{c}\right) - \sum \left(A_{N_{1}}+\frac{1}{5}Z_{c}\right)\right) \nonumber\\
     &=& Z_{c}\left(3.6 A_{i,j}-0.4\sum A_{N_{1}}-1.6Z_{c}\right).\nonumber
\end{eqnarray}
As an example, we detail the energy change
for node $A_{i, j+1}J_{i, j+1}$ belonging to the $N_{2}$ group (the calculation is similar for the other nodes) and obtain
\begin{eqnarray}
    A_{i,j+1}J_{i,j+1} - A'_{i,j+1}J'_{i,j+1} &=&
    \frac{1}{2}A_{i,j+1}\left(4A_{i,j+1} - A_{i, j+2} - A_{i, j} - A_{i+1, j+1} - A_{i-1, j+1}\right) \nonumber\\
    &&-\frac{1}{2}\left(A_{i,j+1} + \frac{1}{5}Z_{c}\right)\left[4\left(A_{i,j+1} + \frac{1}{5}Z_{c}\right) \right.\nonumber\\
    &&- \left.A_{i, j+2} - A_{i, j} + \frac{4}{5}Z_{c}  -A_{i+1, j+1} - A_{i-1, j+1}\right]\nonumber\\
 &=& Z_{c}\left[0.2(A_{i,j}+A_{i,j+2}+A_{i-1,j+1}+A_{i+1,j+1})\right.\nonumber\\ 
    &&\left. -2.4 A_{i,j+1}-0.32Zc\right]. \nonumber
\end{eqnarray}
Computing the energy difference for each node and assembling the results yields:

\begin{eqnarray}
    \Delta E_{AJ} &=& 2.24Z_{c}\left(\frac{1}{2.24}\left[4 A_{i,j}-1.6 \sum_{N_{1}} A_{N_{1}} + 0.2 \sum_{N_{2}} A_{N_{2}}  + 0.4 \sum_{N_{3}} A_{N_{3}}\right]-Zc\right) \nonumber \\ 
    &=&  2.24Z_{c}\left(2 \Delta_{J}A_{i,j} - Z_{c} \right),
\end{eqnarray}
where we have defined
\begin{equation}\label{eq:A.3.}
    \Delta_{J} A_{i,j} = \frac{1}{4.48}\left[4 A_{i,j}-1.6 \sum_{N_{1}} A_{N_{1}} + 0.2 \sum_{N_{2}} A_{N_{2}}  + 0.4 \sum_{N_{3}} A_{N_{3}}\right].
\end{equation}
A positive energy release $\Delta E_{AJ}>0$ therefore requires
\begin{equation}
    \Delta_{J} A_{i,j} > \frac{1}{2}Z_c.
\end{equation}
Again, we can generalize this criterion to $\Delta_{J} A_{i,j} > \alpha Z_{c}$, and for $\alpha = 1$, we recover the quantum release of energy
\begin{equation}\label{eq:e0AJ}
    e_{0} = 2.24Z_{c}\left(2 Z_{c}- Z_{c}\right) = 2.24Z_{c}^2.
\end{equation}
The redistribution must also restore stability. Consider a node $(i,j)$ barely below the instability threshold,
\begin{equation}\label{eq:dum3}
    \frac{1}{4.48}\left[4 A_{i,j}-1.6 \sum_{N_{1}} A_{N_{1}} + 0.2 \sum_{N_{2}} A_{N_{2}}  + 0.4 \sum_{N_{3}} A_{N_{3}}\right] = \alpha Z_{c} - \epsilon,
\end{equation}
receiving a driving increment $\delta A$, and then redistributing. We require that the new curvature be below the instability threshold:
\begin{eqnarray}
    \alpha Z_{c}>\frac{1}{4.48}[&4 \left(A_{i,j} +\delta A- \frac{4}{5}Z_{c}\right)-1.6 \sum_{N_{1}} \left(A_{N_{1}} + \frac{1}{5}Z_{c}\right) + \nonumber\\ &0.2 \sum_{N_{2}} A_{N_{2}}  + 0.4 \sum_{N_{3}} A_{N_{3}}],
\end{eqnarray}
\begin{equation}
    \alpha Z_{c}-\epsilon+ \frac{1}{4.48}\left(4\delta A - 1.6\frac{4}{5}Z_{c} - 4\frac{4}{5}Z_{c}\right)< \alpha Z_{c},
\end{equation}
\begin{equation}
    -\epsilon+ \delta A < 1.12Z_{c}.
\end{equation}
where (\ref{eq:dum3}) was used in the last step.
Even with $\epsilon\ll 1$, stability is always restored provided the increments $\delta A$ is small enough to satisfy:
\begin{equation}
    \delta A < 1.12 Z_{c}.
\end{equation}

\subsection{Characteristic Patterns of Energy Release}\label{app:ModelsEvolution}
\begin{figure}[!h]
    \centering
    \includegraphics[width=12cm]{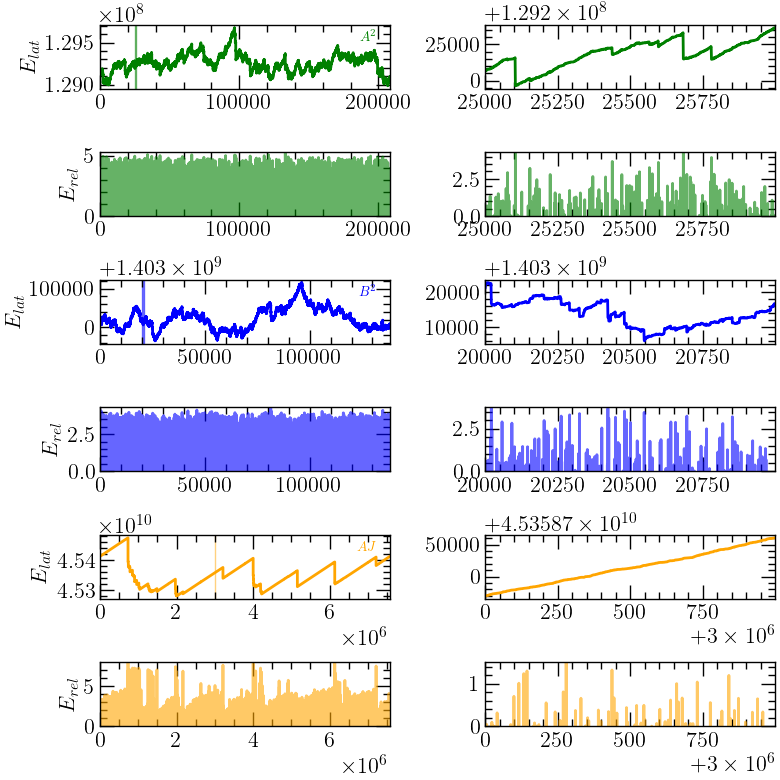}
    \caption{Lattice energy and released energy for the $A^{2}$, $B^{2}$ and $AJ$ models. The number of iterations displayed are different for each model and chosen for clarity of the plots.}
    \label{fig:modelsevo}
\end{figure}

With the new curvature definitions and boundary conditions, under the action of random forcing each model eventually reaches a SOC state. However, these equilibrium states are characterized by distinct dynamics. For example, the $AJ$ model exhibits a unique class of rare, large avalanches.
Figure \ref{fig:modelsevo} displays the characteristic patterns of energy release for each model described in Section \ref{sec:Edefdyn}. The left panels show the evolution of the energy of the lattices and the energy release over an extended time span, whereas the right panels zoom in on the shaded region in the left panels, spanning now only 1000  iterations.
The $A^2$ and $B^2$ models are qualitatively similar, but the $AJ$ model stands distinct from the other two. Nonetheless, all three models exhibit typical SOC behavior in their pattern of intermittent energy release and power-law form of their event size measures.

\section{Analytical Optimization of Energy Release in $A^{2}$ models}\label{app:Analytical} 

We detail in what follows the analytical calculation underlying the redistribution maximizing energy release in the LH model under its conventional $A^{2}$ energy definition (Section \ref{ssec:MCvsF18} and Figure \ref{fig:OptStats}).
We start from the redistribution rules introduced by \cite{Farhangetal18}, as given by Equations \ref{eq:opt1}
in Section \ref{ssec:Opt}.

The optimal $x$ maximizing energy release is calculated via a variational approach, namely by solving d$\Delta E/$d$x = 0$ for $x$:
\begin{eqnarray}
\frac{d}{dx}\Delta E &=& -\frac{d}{dx}(A_{i,j}'^2+A_{i+1,j}'^2+A_{i-1,j}'^2+A_{i,j+1}'^2+A_{i,j-1}'^2)\nonumber\\
0&=&2\left[\left(A_{i+1,j} +\frac{4Z_cr_1}{5(x+a)}\right)\left(\frac{4Z_cr_1}{5(x+a)^2}\right)+
\left(A_{i-1,j} + \frac{4Z_cr_2}{5(x+a)}\right)\left(\frac{4Z_cr_2}{5(x+a)^2}\right)\right.\nonumber\\
&&\left.+\left(A_{i,j+1} + \frac{4Z_cr_3}{5(x+a)}\right)\left(\frac{4Z_cr_3}{5(x+a)^2}\right)-
\left(A_{i,j-1} + \frac{4Z_cx}{5(x+a)}\right)\left(\frac{4Z_ca}{5(x+a)^2}\right)\right] \nonumber\\
0&=&\frac{8Z_c}{5(x+a)^2}\left[r_1\left(A_{i+1,j} +\frac{4Z_cr_1}{5(x+a)}\right)+
r_2\left(A_{i-1,j} +\frac{4Z_cr_2}{5(x+a)}\right)\right.\nonumber\\
&&\left.+r_3\left(A_{i,j+1} + \frac{4Z_cr_3}{5(x+a)}\right)-
a\left(A_{i,j-1} + \frac{4Z_cx}{5(x+a)}\right)\right].\nonumber
\end{eqnarray}
The above equation reduces to
$$0=
\underbrace{r_1A_{i+1,j}+
r_2A_{i-1,j}+
r_3A_{i,j+1} -a
A_{i,j-1}}_{\Theta}+\underbrace{\frac{4Z_c}{5}}_{C}\frac{\overbrace{r_1^2+r_2^2+r_3^2}^{\Phi}-ax}{x+a}$$
$$0=\Theta + \frac{C}{(x+a)}\left(\Phi-ax\right)~,$$
where we have introduced the new variables $\Theta$, $C$, and $\Phi$ that are independent of $x$ to ease the notations. We can now solve for $x$ to obtain
\begin{equation}
\label{eq:x_optimal}
x = \frac{C\Phi + \Theta a}{-\Theta + Ca}\, .
\end{equation}

\section{Power-law Indices}\label{app:Fitting}
\begin{figure}[!h]
    \centering
    \includegraphics[width=12cm]{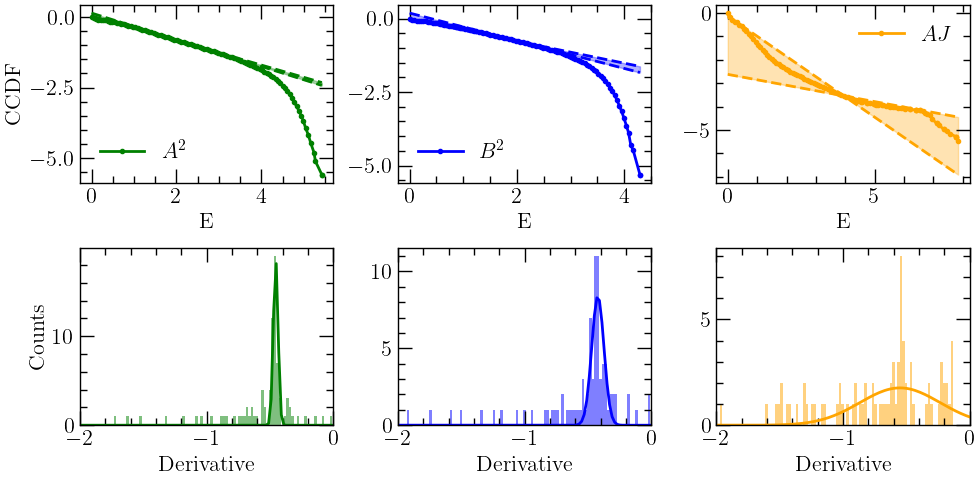}
    \caption{[Top] Complementary cumulative distribution functions (CCDF) for the energy release of each model defined in section \ref{sec:Edefdyn}. A 1$\sigma$ span of $\alpha_{ccdf}$ is shown as an visual guide for each CCDF as a shaded region within dashed lines. [Bottom] Histograms of the slopes of the CCDF and their gaussian best fits are displayed for each model.}
    \label{fig:CCDF}
\end{figure}
\begin{table}[!h]
    \begin{tabular}{c|l|l|l}
        Model & $\alpha_{E}$ & $\alpha_{P}$ & $\alpha_{T}$ \\\hline
        $A^{2}$& $1.43\pm0.02$ & $1.82\pm0.08$ & $1.83\pm0.02$\\
        $B^{2}$ & $1.42\pm0.05$ & $1.8\pm0.2$ & $1.61\pm0.03$\\
        $AJ$ & $1.5\pm0.5$ & $2.0\pm1.0$ & $1.7\pm0.6$\\
        $A^{2}F18$ & $1.35\pm0.06$ & $1.8\pm0.4$ & $3.0\pm1.0$\\
        $A^{2}MC$ & $1.3\pm0.3$ & $1.5\pm0.6$ & $2.1\pm0.2$\\
        $B^{2}MC$ & $1.4\pm0.3$ & $1.4\pm0.05$ & $1.9\pm0.2$\\
        $AJMC$ & $1.4\pm0.2$ & $1.7\pm0.7$ & $1.9\pm0.2$
    \end{tabular}
    \caption{Reported values for the power law slope $\alpha$ of each model presented in this paper.}
    \label{tab:alphas}
\end{table}
Deriving power-law exponents is generally done by specifying arbitrary boundaries for the fitting interval. However, for smoothly varying distributions, the selection of these bounds introduce biases. To alleviate this pitfall, we develop here a method for computing power-law indices without setting arbitrary bounds. We begin by computing the complementary cumulative distribution function (CCDF). This function is generally smoother than the PDF since the latter requires binning of the data. The power-law slope for the CCDF is related to the power-law slope of the PDF by:
\begin{equation}\label{eq:CCDF}
    \alpha = 1-\alpha_{ccdf}\, ,
\end{equation}
where PDF$(X) = X^{\alpha}$ for any quantity $X$. In order to find $\alpha_{ccdf}$, we  sample the slope of the CCDF along a logarithmic scale  to avoid giving too much weight to zones with higher point density. We then create a histogram of the slope values and fit that distribution with a Gaussian function, as shown in Figure \ref{fig:CCDF} for $E$ of models $A^2$, $B^2$ and $AJ$. In order to reduce the uncertainty caused by the tail of the distribution, the Gaussian function is recursively fitted by ignoring bins outside 3$\sigma$ of the mean until the mean converges within 0.01\%. The value and uncertainty for $\alpha_{ccdf}$ are then reported as the mean and width of the Gaussian. This method attributes large uncertainties for distributions with significant deviations from a singular power-law, such as the $AJ$ model in Figure \ref{fig:EnergyStats}, panel C. The values for $\alpha$ for each model, with associated error bars, are listed in Table \ref{tab:alphas}.


\begin{acks}
We wish to thank Nastaran Farhang for very useful exchanges on the inner workings of the
\cite{Farhangetal18} avalanche model. We also thank Christian Thibeault for useful discussions.
\end{acks}

\begin{fundinginformation}
This research was supported by
NSERC Discovery grant RGPIN/05278-2018 (PC), and a merit scholarship from the Universit\'e de Montr\'eal's Physics department (HL). AS acknowledges support from ANR STORMGENESIS \#ANR-22-CE31-0013-01, DIM-ACAV+ ANAIS2 project, ERC Whole Sun Synergy grant \#810218, INSU/PNST, and Solar Orbiter CNES funds.
\end{fundinginformation}

%
%
\bibliographystyle{spr-mp-sola}
\typeout{}
\bibliography{Bibliography}  
%
%
%
%
\end{appendix}
\end{article} 
\end{document}